\documentclass[twocolumn,preprintnumbers,amsmath,amssymb]{revtex4}

\usepackage{amssymb}
\usepackage{graphicx}
\usepackage{dcolumn}
\usepackage{amsmath}
\usepackage{bm}
\usepackage[latin1]{inputenc}
\newcommand{\comment}[1]{}
\newcommand{\ud}{\mathrm{d}}

\begin{document}




\title{Reynolds number effect on the velocity increment skewness in isotropic turbulence}


\author{Wouter J.T. Bos$^1$, Laurent Chevillard$^2$, Julian
  F. Scott$^1$ and Robert Rubinstein$^3$}

\affiliation{$^1$ LMFA, CNRS, Ecole Centrale de Lyon, Universit\'e de Lyon, 
  69134 Ecully, France\\
$^2$ Laboratoire de Physique, ENS Lyon, CNRS, Universit\'e de Lyon, Lyon, France\\
$^3$ Newport News, VA, USA}

\begin{abstract}
Second and third order longitudinal structure functions and wavenumber spectra of isotropic turbulence are computed using the EDQNM model and compared to results of the multifractal formalism. At the highest Reynolds number available in windtunnel experiments, $R_\lambda=2500$, both the multifractal model and EDQNM give power-law corrections to the inertial range scaling of the velocity increment skewness. For EDQNM, this correction is a finite Reynolds number effect, whereas for the multifractal formalism it is an intermittency correction that persists at any high Reynolds number. Furthermore, the two approaches yield realistic behavior of second and third order statistics of the velocity fluctuations in the dissipative and near-dissipative ranges. Similarities and differences are highlighted, in particular the Reynolds number dependence. 
\end{abstract}


\maketitle


\section{Introduction}

The nonlinearity in the Navier-Stokes equations gives rise to an
interaction between different length-scales in a turbulent flow. These
interactions are the basic mechanism behind the celebrated
Kolmogorov-Richardson energy cascade  \cite{Kolmogorov,Richardson}. This phenomenological picture of energy cascading from scale to scale towards the scales in which dissipation becomes appreciable is the cornerstone of a large number of turbulence models (\emph{e.g.} reference \cite{Lumley92}). If locality in scale-space is assumed, energy-conservation and local-isotropy will lead to a wavenumber dependence of the energy spectrum of the form
\begin{equation}\label{K41}
E(k)\sim \epsilon^{2/3}k^{-5/3}
\end{equation}
with $\epsilon$ the energy flux, which, using the assumption of statistical stationarity, equals the energy dissipation. A physical space equivalent of this scaling law is the scale dependence of the second-order longitudinal structure function,
\begin{equation}
D_{LL}(r)\sim \epsilon^{2/3}r^{2/3}.
\end{equation}
The definitions of $D_{LL}(r)$ and $E(k)$  will be given below.

The possibility of corrections to the inertial range scaling of
structure functions, due to the intermittent character of the energy
dissipation \cite{LandauBook}, was taken into account in a more general theory advanced by Kolmogorov and Oboukhov 
\cite{Kolmogorov1962,Obou62}. Experiments aiming at the measurement of the intermittency corrections 
(\emph{e.g.} reference \cite{Atta1970,Anselmet1984}) indeed showed small corrections to the scaling which could be due 
to intermittency, in particular for higher-order structure functions. 
Subsequently a large number of phenomenological models was proposed to describe the intermittent character of turbulence. 
Reference \cite{FrischBook} gives an overview of work on intermittency upto 1995. One of the more successful models, 
in the sense of reproducing the different features of isotropic turbulence, is the multifractal model \cite{FrischBook}. 
This phenomenological description  compares well to measurements and gives non-zero intermittency corrections 
to the inertial range scaling of the energy spectrum and of higher order quantities.

A valuable theoretical tool to study the statistical properties of homogeneous turbulence is two-point closure theory. 
The first theoretical approach of this kind, derived from the Navier-Stokes equations, is the Direct Interaction Approximation 
(DIA) \cite{KraichnanDIA}. Subsequent improvements \cite{Kraichnan65} of this theory allowed to show that the 
$k^{-5/3}$ dependence of the energy spectrum can be related directly to the Navier-Stokes equations. 
Simplifications led to different related closures such as the test-field model \cite{KraichnanTFM}, the Lagrangian 
renormalized approximation \cite{Kaneda81} and the Eddy-Damped Quasi-Normal Approximation (EDQNM) \cite{Orszag}. 
EDQNM is of the closures named here the simplest. It is obtained by assuming in the DIA formulation that the 
two-time correlations decay exponentially with a typical time-scale modeled phenomenologically. We note that 
this time-scale can also be determined self-consistently within the EDQNM approach \cite{Bos2006}.

These closures, although directly related to the Navier-Stokes equations, do not yield any intermittency corrections to the scaling 
(\ref{K41}). However, the predicted results for scaling
exponents of the energy spectrum compare rather well to experimentally 
observed values \cite{Bos2005}. 
Indeed, at low Reynolds numbers, corrections to the scaling exponents due to the finite Reynolds number are usually larger 
than the expected intermittency corrections and these finite Reynolds number effects vanish very slowly \cite{Qian1997,Qian1999,Bos2005}. 
The fact that two-point closure and the multifractal formalism can treat both low and very high Reynolds 
numbers using limited computational effort, makes these approaches very attractive to study Reynolds-number effects.

The present work will compare the predictions of closure for second
and third-order quantities with results of the multifractal description. This
will allow to show to what extent intermittency corrections can be
distinguished from Reynolds number effects at low, moderate and high
Reynolds numbers. We choose to compare quantities in physical space,
since most experimental and theoretical efforts aiming at the understanding and description of intermittency focus on these 
quantities (we note however that in principle intermittency
corrections, if any, should also be observed in wavenumber spectra). Therefore we need to convert the Fourier-space quantities into physical space quantities. 
It is described in the next section how this is done. The relations
to convert physical space quantities into their Fourier-space 
counterparts is also given.  
In section \ref{sec3} we will present the EDQNM model and we will give
an outline of the multifractal description. 
In section \ref{sec4} we present the results of the EDQNM model for
these quantities and compare with the multifractal results in both
Fourier and physical space. 
Section \ref{sec5} concludes this article.

\section{Exact relations between second and third order quantities in Fourier space and in physical space}\label{sec2}

In this section we will give the relation between the energy spectrum
and the second order structure function $D_{LL}(r)$, and between the
nonlinear transfer and the third-order longitudinal velocity structure
function, $D_{LLL}(r)$. Even though the relations given here are not
new (e.g. \cite{BookBatchelor,Monin,MathieuBook}), the details of the
derivation are dispersed or not well documented in literature and we
think that it is therefore worth to write down in detail this
derivation, which can be found in the appendix.

\subsection{Derivation of the Lin-equation}

The starting point is the Navier-Stokes equations for incompressible flow,
\begin{eqnarray}
\frac{\partial u_i(\bm x)}{\partial t}+u_j(\bm x)\frac{\partial u_i(\bm x)}{\partial x_j}=-\frac{1}{\rho}\frac{\partial p(\bm x)}{\partial x_i}+\nu \frac{\partial^2 u_i(\bm x)}{\partial x_j^2}\\
\frac{\partial u_i(\bm x)}{\partial x_i}=0,
\end{eqnarray}
with $\rho$ the density and $p$ the pressure. Time arguments are
omitted for brevity. The three-dimensional Fourier transfer is defined as
\begin{equation}
u_i(\bm k)=\frac{1}{(2\pi)^{3}}\int u_i(\bm x) e^{-i\bm k \cdot \bm r}d\bm k.
\end{equation}
In Fourier space the Navier-Stokes equations can be written
\begin{eqnarray}\label{NSF}
\frac{\partial u_i(\bm k)}{\partial t} +\nu k^2 u_i(\bm k)=\nonumber\\
-\frac{i}{2}P_{ijm}(\bm k)\iint u_j(\bm p)u_m(\bm q)\delta(\bm k -\bm p-\bm q)\ud\bm p\ud\bm q
\end{eqnarray}
with
\begin{eqnarray}
P_{ijm}(\bm k)=k_jP_{im}(\bm k)+k_mP_{ij}(\bm k),\\
P_{ij}(\bm k)=\delta_{ij}-\frac{k_ik_j}{k^2}.
\end{eqnarray}
To derive (\ref{NSF}), the incompressibility condition was used to eliminate the pressure term. In isotropic non-helical turbulence, the energy spectrum is related to $u_i(\bm k)$ by
\begin{eqnarray}\label{eqEk}
\frac{P_{ij}(\bm k)}{4\pi k^2}E(k)=\overline{u_i(\bm k)u_j^*(\bm k)}
\end{eqnarray}
and since $u_i(\bm x)$ is real, this gives
\begin{eqnarray}
E(k)=2\pi k^2\overline{u_i(\bm k)u_i(-\bm k)}.
\end{eqnarray}
In order to derive the equation for $E(k)$, we multiply (\ref{NSF}) by $u_i(-\bm k)$. Then we write a similar equation for $u_i(-\bm k)$ and multiply by $u_i(\bm k)$ Summing both equations and averaging yields,
\begin{eqnarray}\label{Lin2}
\left[\frac{\partial}{\partial t}+2\nu k^2\right]E(k)&=&i\pi k^2 P_{ijm}(\bm k)\left[T_{ijm}(\bm k)-T_{ijm}^*(\bm k)\right]\nonumber\\
&=& T(k)
\end{eqnarray}
with
\begin{eqnarray}\label{T1}
T_{ijm}(\bm k)=\iint \overline{u_i(\bm k)u_j(\bm p)u_m(\bm q)}\delta(\bm k +\bm p+\bm q)\ud\bm p\ud\bm q \nonumber\\
\\
T_{ijm}^*(\bm k)=\iint \overline{u_i(-\bm k)u_j(-\bm p)u_m(-\bm q)}\delta(\bm k +\bm p+\bm q)\ud\bm p\ud\bm q. \nonumber\\
\label{T2}
\end{eqnarray}
By isotropy it can be shown that $T_{ijm}^*(\bm k)=-T_{ijm}(\bm
k)$. The RHS of the Lin-equation (\ref{Lin2}) is the nonlinear transfer $T(k)$, which we will relate to the third-order longitudinal structure function. But first we will give the relation between the energy spectrum and the second order longitudinal structure function. 

\subsection{Relation between $E(k)$ and $D_{LL}(r)$.}

The second order longitudinal structure function is defined as
\begin{equation}\label{eqDLL0}
D_{LL}(r)=\overline{\delta u_L^2}
\end{equation}
with 
\begin{equation}
\delta u_L=u_L-u'_L=\frac{r_i}{r}u_i(\bm x)-\frac{r_i}{r}u_i(\bm x+\bm r).
\end{equation}
its relation to the energy spectrum is 
\begin{eqnarray}\label{eqDllF}
D_{LL}(r)&=&\int E(k)f(kr) \ud k
\end{eqnarray}
with $f(x)$ given by
\begin{eqnarray}\label{fx}
f(x)&=&4\left[ \frac{1}{3}- \frac{\sin (x) -(x)\cos (x)}{(x)^3}.
  \right]
\end{eqnarray}
The derivation of this expression is given in the appendix. A
convenient expression to compute the energy spectrum from the second
order structure function is 
\begin{eqnarray}\label{DtoE}
E(k)=\frac{\overline{u^2}}{\pi}\int
\left(1-\frac{D_{LL}(r)}{2\overline{u^2}}\right)kr \left[\sin(kr) -kr \cos(kr)\right] \ud r\nonumber\\
\end{eqnarray}
and we refer to Mathieu and Scott \cite{MathieuBook} for the derivation.

\subsection{Relation between $T(k)$ and $D_{LLL}(r)$.}

The third order longitudinal structure function in homogeneous
turbulence can be expressed as
\begin{equation}
D_{LLL}(r)=\overline{\delta u_L^3}=3\left(\overline{u_L{u'_L}^2}-\overline{u'_Lu_L^2} \right).
\end{equation}
which is related to the transfer spectrum by
\begin{eqnarray}\label{eqDlll5}
D_{LLL}(r)= r\int_0^{\infty} T(k) g(kr)\ud k,
\end{eqnarray}
with 
\begin{equation}\label{gx}
g(x)=12\frac{3 \left(\sin x-x\cos x\right)-x^2\sin x}{x^5},
\end{equation}
with details given in the appendix. The equivalent expression to
compute the transfer spectrum from $D_{LLL}(r)$ is \cite{MathieuBook},
\begin{equation}\label{DtoT}
T(k)=\frac{k}{6\pi}\int \frac{\sin(kr)}{r}\frac{\partial }{\partial
  r}\left[\frac{1}{r}\frac{\partial}{\partial r}\left(r^4 D_{LLL}(r) \right) \right]       \ud r.
\end{equation}

\subsection{Small scale behavior of $D_{LL}(r)$ and $D_{LLL}(r)$}

Before continuing, let us have a look at the behavior of the functions (\ref{fx}) and (\ref{gx}),
\begin{eqnarray}
f(x)=4\left[ \frac{1}{3}- \frac{\sin x -x\cos x}{x^3} \right]\\
g(x)=12 \frac{3 \left(\sin x-x\cos x\right)-x^2\sin x}{x^5}.
\end{eqnarray}
Taylor expansions of the sine and cosine terms show that for $x\downarrow 0$,
\begin{eqnarray}
f(x)=\frac{2}{15}x^2 + \mathcal O(x^3)\\
g(x)=\frac{4}{5}-\frac{2}{35}x^2+\mathcal O(x^4).
\end{eqnarray}
Using this in (\ref{eqDllF}) and (\ref{eqDlll5}), we find for very small $r$,
\begin{eqnarray}
D_{LL}(r)&=&\frac{2}{15}r^2\int k^2 E(k) \ud k\label{dll00}\nonumber\\
&=&\frac{\epsilon r^2}{15\nu}\label{dll0}\\
D_{LLL}(r)&=&\frac{4}{5}r\int T(k) \ud k-\frac{2}{35}r^3\int k^2 T(k) \ud k\nonumber\\
&=&-\frac{2}{35}r^3\int k^2 T(k) \ud k\label{dlll0}
\end{eqnarray}
in which we used that
\begin{eqnarray}\label{defeps}
2\nu\int k^2 E(k)\ud k&=&\epsilon\\
\int T(k) \ud k &=&0,
\end{eqnarray}
with $\epsilon$ the energy dissipation. So we find that the structure functions of order $2$ and $3$ scale as $r^2$ and $r^3$ 
respectively for very small $r$, which is expected since at small enough scales the flow can be considered as smooth.

The velocity-increment skewness is defined as 
\begin{eqnarray}
S(r)=\frac{D_{LLL}(r)}{D_{LL}(r)^{3/2}}.
\end{eqnarray}
Since at very small scales 
\begin{eqnarray}
\delta u_L\approx r\frac{\partial u}{\partial x},
\end{eqnarray}
one finds that
\begin{eqnarray}
\lim_{r\to 0} S(r)=\frac{\overline{(\partial u/\partial x)^3}}{\left(\overline{(\partial u/\partial x)^2}\right)^{3/2}}.
\end{eqnarray}
Using expressions (\ref{dll00}) and (\ref{dlll0}), we find \cite{BookBatchelor},
\begin{eqnarray}\label{skew0}
\lim_{r\to 0}
S(r)=\frac{[(\partial_xu)^3]}{[(\partial_xu)^2]^{3/2}}=-\frac{15^{3/2}}{35(2)^{1/2}}\frac{\int_0^{\infty}k^2 T(k)dk}{[
\int_0^{\infty}k^2 E(k)dk
]^{3/2}}.
\end{eqnarray}
In the case of high-Reynolds number, if the non-stationarity can be neglected at high $k$, or if the turbulence is kept stationary by a forcing term acting only at small $k$, we have
\begin{eqnarray}
\int k^2 T(k) \ud k \approx \int 2\nu k^4 E(k) \ud k,
\end{eqnarray}
so that \cite{BookBatchelor}
\begin{eqnarray}\label{SBatch}
\lim_{r\to 0}
S(r)\approx-\frac{15^{3/2}2^{1/2}\nu}{35}\frac{\int_0^{\infty}k^4 E(k)dk}{[
\int_0^{\infty}k^2 E(k)dk
]^{3/2}}.
\end{eqnarray}
The velocity-derivative skewness is then completely determined by
moments of the energy spectrum. Using (\ref{defeps}), and assuming an
inertial range spectrum  extending upto $k_f$ of the form
\begin{equation}
E(k)\sim \epsilon^{2/3}k^a
\end{equation}
we obtain for $\alpha >-3$
\begin{equation}
k_f\sim\left(\epsilon^{1/3}/\nu\right)^{1/(\alpha+3)}.
\end{equation}
Substituting this in (\ref{SBatch}) we obtain,
\begin{eqnarray}
\lim_{r\to 0}
S(r)&\sim& k_f^{-\frac{1}{2}(5+3\alpha)}\nonumber\\
&\sim& R_\lambda^{-\frac{3\alpha+5}{\alpha+3}},\label{SkewRe}
\end{eqnarray}
in which we used that the Taylor-scale Reynolds number, to be defined
later, is proportional to $\nu^{-1/2}$. It follows from this expression that the velocity derivative skewness is independent of the
Reynolds number if $\alpha= -5/3$. If corrections to the
Kolmogorov 1941 (K41) scaling are present, as is the case in the multi-fractal
formalism, the skewness becomes a function of the Reynolds number. This dependence is by (\ref{SkewRe}) directly related to the intermittency correction to the K41 scaling.

\subsection{Large scale behavior of $D_{LL}(r)$ and $D_{LLL}(r)$}

At large $r$ we find
\begin{eqnarray}
D_{LL}(r)&=&\frac{4}{3}\int E(k) \ud k=2\overline{u^2}
\end{eqnarray}
which is expected from (\ref{eqDll0}) since at large separation
distances the correlation between the velocity at two points is
supposed to vanish. $D_{LLL}(r)$ tends for the same reason to zero for
large separation distances $r$. At large $r$ the velocity increment
skewness should therefore go smoothly to zero, since the velocity
correlation should decay smoothly at large $r$. The exact way in which
$D_{LLL}(r)$ tends to zero depends on the behavior of the energy
spectrum at the very low wavenumbers.

\section{The EDQNM model and the multifractal description \label{sec3}}

\subsection{The EDQNM model}

The EDQNM model is a closure of the Lin-equation in which the
nonlinear transfer $T(k)$ is expressed as a function of the energy
spectrum. The transfer $T(k)$ is given by 
\begin{eqnarray}\label{TNL}
T(k)=\iint_{\Delta}\Theta_{kpq}~(xy+z^3)\left[k^2p~E(p)E(q) \right.\nonumber\\
\left. -p^3E(q)E(k)\right]\frac{\ud p\ud q}{pq}.~~~~~~~~~
\end{eqnarray}
In equation (\ref{TNL}), $\Delta$ is a band in $p,q$-space so that the
three wave-vectors ${\bm{k}, \bm{p}, \bm{q}}$ form a triangle. $x,y,z$
are the cosines of the angles opposite to ${\bm{k}, \bm{p}, \bm{q}}$ in this
triangle. This particular structure is common to all closures derived
from the Direct Interaction Approximation \cite{KraichnanDIA}.
DIA is a self-consistent two-point two-time theory without adjustable
parameters. Simplifications are needed to obtain a single-time (or
Markovian) description introducting assumptions and
adjustable parameters. In the case of EDQNM the simplifying assumption
is that all time-correlations decay exponentially, with a time-scale
$\Theta_{kpq}$ modeled phenomenologically by
\begin{equation}\label{eqTheta}
\Theta_{kpq}=\frac{1-\exp(-(\eta_k+\eta_p+\eta_q)\times t)}{\eta_k+\eta_p+\eta_q}
\end{equation}
in which $\eta$ is the eddy damping, expressed as
\begin{equation}
\eta_k=\lambda\sqrt{\int_0^k s^2E(s)\ud s} +\nu k^2.
\end{equation}
related to the timescale associated with an eddy at wavenumber $k$,
parameterised by the EDQNM parameter, $\lambda$, which is chosen equal
to $0.49$ \cite{Bos2006}. The exponential time-dependence in
(\ref{eqTheta}) appears by the assumption
that the initial conditions have vanishing triple correlation  as
would be the case for a Gaussian field. Its influence 
 vanishes at long times. For an extensive discussion of the EDQNM model see \cite{Lesieur,SagautBook}, but we want to stress that one of the key features of EDQNM is that it is applicable at all Reynolds numbers (it is not an asymptotic theory) and at all scales of a turbulent flow. In other words, its results go beyond mere scaling and can give insights on the Reynolds number dependency of different quantities related to turbulence.

We performed simulations of the EDQNM model in the unforced
case by integrating numerically Eq.~(\ref{TNL}), starting from an initial spectrum,
\begin{equation} \label{inicond}
E_k(0)=B k^4\exp\left[-(k/k_L)^2\right],
\end{equation}
with $B$ chosen to normalize the energy to unity and $k_L=8 k_0$, $k_0$ being the smallest wavenumber. The resolution is chosen $12$ gridpoints per decade, logarithmically spaced. In the decaying simulations results are evaluated in the self-similar stage of decay, in which $\epsilon/e_{\textrm{kin}}$, with $e_{\textrm{kin}}$ the kinetic energy, is proportional to $t^{-1}$. Forced simulations are evaluated when a steady state is reached. The forcing in these cases corresponds to a region $k<k_L$, in which the energy spectrum is kept constant.

\subsection{The multifractal description}

In this section some key concepts of the multi-fractal description
will be presented. A more detailed presentation and references can be found in appendix \ref{AppMF}.

In the multifractal description the velocity-increments $\delta
u_L(\bm x,r) = u_L(\bm x+r)-u_L(\bm x)$ at scales $r$ in the inertial range are modeled by the product of two independent random variables 
\begin{equation}
\delta u_L(x,r) = \beta_r \xi.
\end{equation}
In this expression $\xi$ is a zero average Gaussian random variable of variance $\sigma^2$, where $\sigma^2$ is twice the mean-square of the velocity fluctuations. The quantity $\beta_r$ introduces the scale dependence in the statistics of the velocity increments. It is defined as 
\begin{equation}
\beta_r = \left(\frac{r}{L_0}\right)^h.
\end{equation}
with $L_0$ the integral lengthscale. The particularity of the approach
lies in the fact that the exponent $h$ is a fluctuating quantity. If
a constant value $h=1/3$ is chosen, K41 behavior is recovered. In the
multifractal framework $h$ is determined by the probability density function 
\begin{equation}
\mathcal P_r(h) \propto \left(\frac{r}{L_0}\right)^{1-\mathcal D(h)}.
\end{equation}
If the unknown function $\mathcal D(h)$ is given (for example by
comparison with experimental results), a complete description of the
inertial range statistics of the velocity increments can be
obtained. An extension to take into account the dissipative effects
was proposed by Paladin and Vulpiani \cite{PalVul87} and Nelkin
\cite{Nel90}. Further details on the multi-fractal description,
including the expressions for $\beta_r$ and $P_r(h)$, are given in the appendix. 

\begin{figure}
\centering
\setlength{\unitlength}{1\textwidth}
\includegraphics[width=0.45\unitlength]{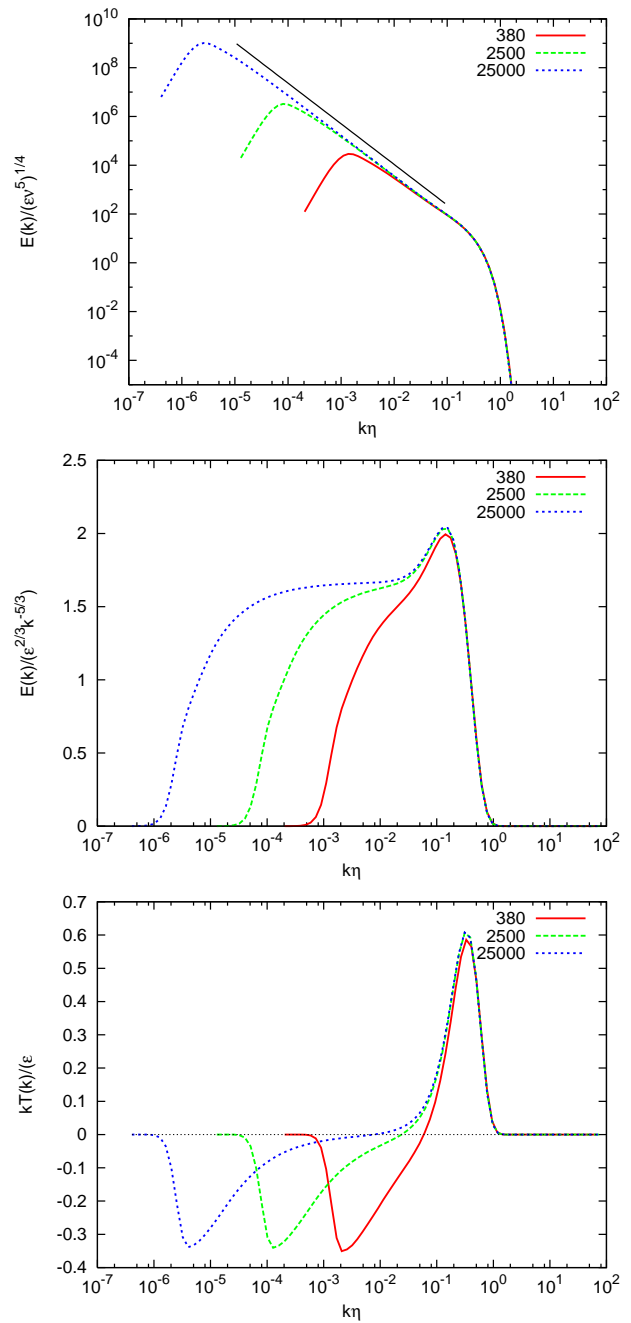}
\caption{Results for the energy spectrum computed by the EDQNM model. In the center plot we show the compensated spectrum. In the bottom figure the nonlinear transfer is plotted. All quantities are normalized by Kolmogorov scales. \label{FigEk}}
\end{figure}

\begin{figure}
\centering
\setlength{\unitlength}{0.45\textwidth}
\includegraphics[width=1\unitlength]{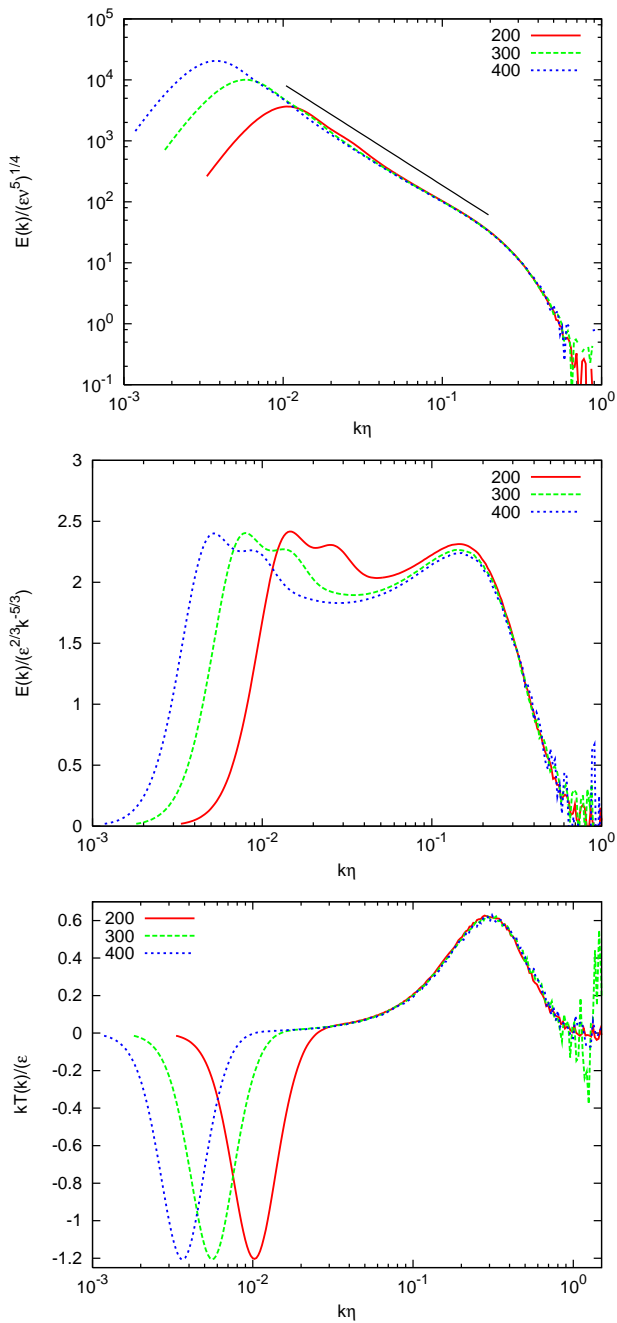}
\caption{Results for the energy spectrum computed from the
  multi-fractal description. In the center plot we show the compensated spectrum. In the bottom figure the nonlinear transfer is plotted. \label{FigEkMF}}
\end{figure}

\section{Results for second and third order quantities}\label{sec4}

\subsection{Results in Fourier space}

In the following we will consider three different values of the Reynolds number
\begin{eqnarray}
R_\lambda=\sqrt{15\frac{\overline{u^2}^2}{\nu \epsilon}}.
\end{eqnarray}
These values are $R_\lambda=380,~2500,~25000$. The lowest corresponds
to a typical Reynolds number for laboratory experiments in jets or
wind-tunnels, the second one to the highest Reynolds number obtained
in wind-tunnel turbulence, \emph{i.e.} in the Modane windtunnel
\cite{Kahalerras1998}, and $R_\lambda=25000$ corresponds to the
Reynolds number of large scale atmospheric flows and no controlled
experimental results of isotropic turbulence are available. In the
following we will present results for these Reynolds numbers. All
quantities are normalized by Kolmogorov scales, which means that they
are non-dimensionalized by using the variables $\nu$ and
$\epsilon$. For example, all lengthscales are normalized by
$\eta=(\nu^3/\epsilon)^{1/4}$. This normalization allows to collapse
the dissipation range of the different quantities if this range
becomes independent of the viscosity. This is the case in the
K41 phenomenology. In the presence of intermittency this is not the case anymore. It will however been shown in the following that also in that case the dissipation ranges of the different quantities nearly collapse in the present range of Reynolds numbers.

In Figure \ref{FigEk}, the energy spectrum is shown for three distinct
Taylor-scale Reynolds numbers. We observe a clear $k^{-5/3}$ power-law
in the log-log representation. However, when showing the compensated
spectra in log-lin representation it is observed that only at the
highest Reynolds number a clear plateau can be discerned. At small $k$
this plateau drops to zero, and at large $k$ a viscous bottleneck is
observed.

Since we are interested in second and third order quantities in the present work, we also show the nonlinear transfer. Again we observe that the asymptotic case, here indicated by a plateau around zero in between the negative and the positive lobe of the transfer spectrum, is only observed at the highest Reynolds number.

The energy and transfer spectra computed from the multifractal
description are shown in Fig. \ref{FigEkMF}. The results are shown for
relatively low Reynolds numbers (upto $R_\lambda=400$), since the
numerical integration for higher values yielded extremely noisy
results in the dissipation range. A bumpy large-scale behavior is
observed in the compensated energy spectra, corresponding to the
\emph{ad-hoc} modeling of the large scales, as explained in appendix \ref{AppMF}. This modeling also causes
the relatively narrow negative peak in the transfer spectrum. In the
following, when presenting the structure functions, we do not need a
smooth behavior for the large-scales and we will therefore not use the
\emph{ad-hoc} modification of the large scales. In the dissipative and near-dissipative ranges, the spectral quantities (such as power spectrum and nonlinear transfer) obtained from EDQNM closures and the multifractal formalism 
are very similar. In the dissipation range a viscous bottleneck is
observed in both descriptions.

\begin{figure}
\centering
\setlength{\unitlength}{0.45\textwidth}
\includegraphics[width=1\unitlength]{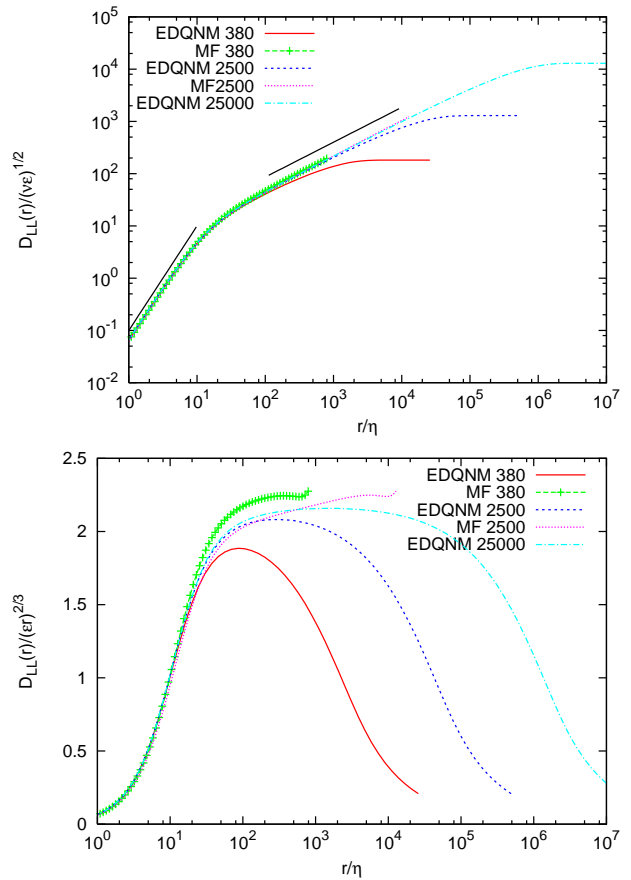}
\caption{The second order londitudinal structure function computed by EDQNM and the multifractal model. Straight black lines indicate powerlaw behavior proportional to $r^2$ and $r^{2/3}$ respectively. In the bottom figure the functions are compensated according to K41 scaling. \label{figsf2}}
\end{figure}

\begin{figure}
\centering
\setlength{\unitlength}{0.45\textwidth}
\includegraphics[width=1\unitlength]{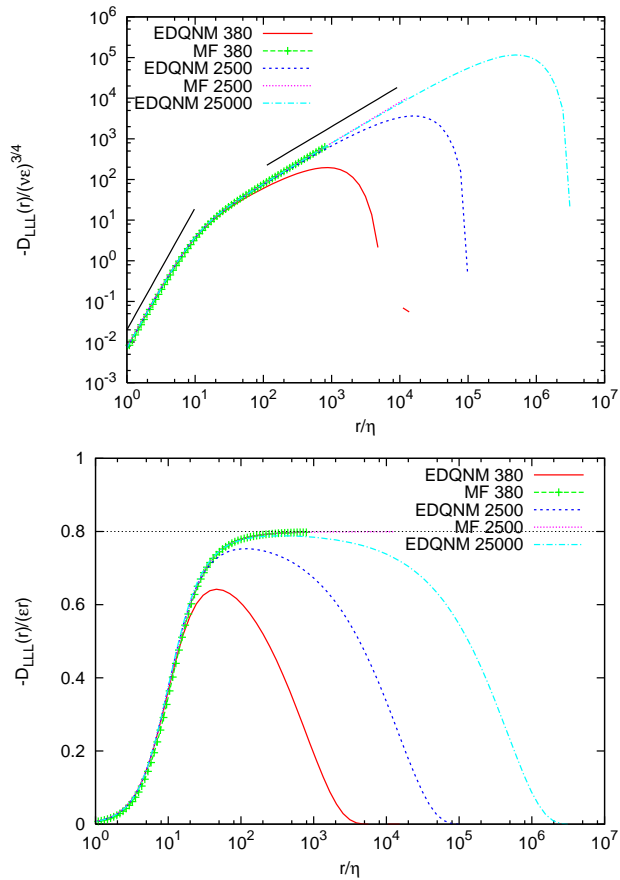}
\caption{The third order londitudinal structure function computed by EDQNM and the multifractal model. Straight black lines indicate powerlaw behavior proportional to $r^3$ and $r^{1}$ respectively. In the bottom figure the functions are compensated according to K41 scaling. The asymptotic result $4/5$ is indicated by a dashed line. \label{figsf3}}
\end{figure}

\subsection{Results for structure functions}

We used equations (\ref{eqDllF}) and (\ref{eqDlll5}) to compute the structure functions from the energy spectra and transfer spectra shown in the previous section. The results for the second order structure function are shown in figure \ref{figsf2}. We show the multifractal prediction in the same graph. In the log-log representation we clearly observe the smooth $r^2$ small scale behavior and the plateau proportional to the kinetic energy at large scales. In between a power-law dependence close to $r^{2/3}$ is observed. 

The multifractal prediction closely ressembles the EDQNM result in the
dissipation range. The differences between the two models are more
clearly visible in the compensated plot, where we observe that for
$R_\lambda=2500$, the power-law dependence is clearly steeper than
$r^{2/3}$. The largest difference is observed at large $r$. Indeed,
the multifractal description does not take into account the shape of
the velocity correlation at large $r$. This correlation should in a
realistic flow smoothly tend to zero, but this effect is not taken
into account in the formalism. Note that we prefer to show here the results
without the \emph{ad-hoc} modification proposed in the last section. We further observe that the structure function computed from EDQNM, as for the energy spectra, does not display a clear plateau in the compensated representation for $R_\lambda<25000$.

The results for the third-order structure functions are shown in figure \ref{figsf3}. Again we clearly observe the smooth small-scale behavior proportional to $r^3$. In this range the multifractal model closely follows the EDQNM results. For larger $r$ a close to linear dependence and at large scales a decrease towards zero. Also here the multifractal formalism does not take into account the large scales. In the inertial range, at very large $R_\lambda$, the third order structure function should scale as
\begin{equation}\label{45th}
D_{LLL}(r)=-\frac{4}{5}\epsilon r.
\end{equation}
it is observed that this is only reached at the highest $R_\lambda$ for EDQNM and only for a short range of scales. The multi-fractal results collapse with (\ref{45th}) already at $R_\lambda=380$.

\begin{figure}
\centering
\setlength{\unitlength}{0.45\textwidth}
\includegraphics[width=1\unitlength]{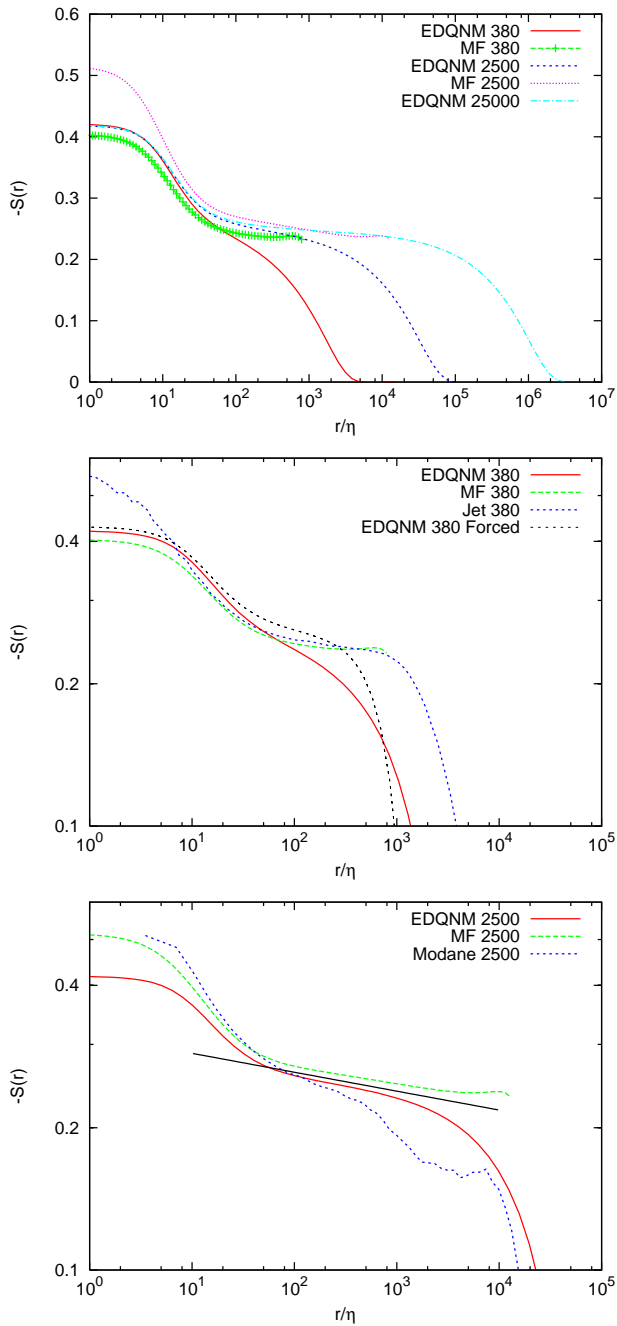}
\caption{Comparison of the longitudinal velocity increment skewness between EDQNM and the multifractal approach. In the center figure results are compared to air-jet results. In the bottom plot the results are compared to high-Reynolds number wind-tunnel experiments. The straight black line indicates a powerlaw proportional to $r^{-0.04}$\label{skew}}
\end{figure}

\subsection{Results for the velocity increment and derivative skewness}

In figure \ref{skew} top, we show the velocity increment skewness for different Reynolds numbers. 
In the K41 phenomenology, this quantity should give a constant value in the inertial range. 
It is observed that the fact that $D_{LLL}(r)$ tends to zero smoothly for large $r$ results in a gently decreasing function, 
rather than a constant value.

\begin{figure}
\centering
\setlength{\unitlength}{0.45\textwidth}
\includegraphics[width=1\unitlength]{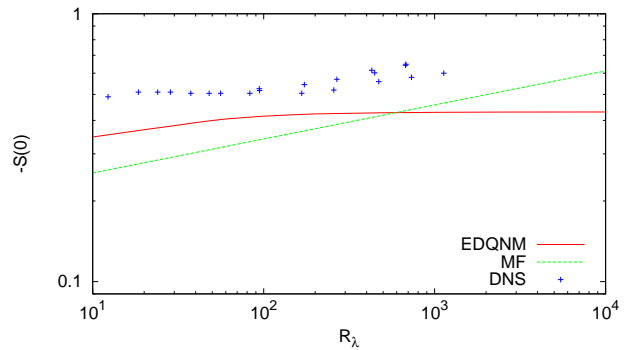}
\caption{The velocity derivative skewness as a function of the Reynolds numbers for EDQNM and MF. Also shown are DNS results from references \cite{Ishihara2007,Kerr1985}.
\label{skewRe}}
\end{figure}

In the dissipation range all curves nearly collapse. Only the multifractal approach gives a slightly higher value than the rest, since the velocity derivative skewness is a function of $R_\lambda$, as will be shown later, in figure \ref{skewRe}. At large scales, the multifractal result closely follows the high-Reynolds EDQNM result up to the cut-off of the multi-fractal result.

In the center and top graph of figure \ref{skew}, we compare the
results also with experimental results. At $R_\lambda=380$, we compare
with hot-wire measurements in an air-jet experiment \cite{Chevillard2006-2}.  At small scales the experimental value is significantly larger than the theoretical results. At these scales the accuracy of the hot-wire probe decreases however. In the inertial range the multi-fractal approach is very close to the experimental value. The EDQNM curve drops much faster to zero. Inhomogeneity and anisotropy of the experimental turbulent field could be behind this discrepancy.

At $R_\lambda=2500$ a comparison is made with the velocity increment skewness measured in the return-channel of the ONERA wind-tunnel in Modane. The Reynolds number obtained there is one of the highest measured in wind-tunnel turbulence. Unfortunately at large scales the third-order statistics are not fully converged so that no smooth curve is available there. However the general trend of the curves is quite similar at all scales. A surprising fact is here the power-law that is observed in the inertial range of both the multifractal result and the EDQNM result. Indeed, in the multifractal approach this power-law is a signature of inertial-range intermittency and the model is developed to take this into account. In the EDQNM approach, however, this power-law is a transient effect, due to the finite-Reynolds number. In EDQNM this power-law vanishes thus at high Reynolds number.
 
In figure \ref{skewRe} we show the Reynolds number dependency of the longitudinal velocity derivative skewness, 
as computed by equation (\ref{skew0}). We see that this quantity saturates for $R_\lambda>100$ at a value around $0.4$. In the multifractal approach this quantity follows a power-law of the Reynolds number (see Eqs. (\ref{eq:pred1}) and 
(\ref{eq:predSkewGradLN})) with an exponent around $0.13$. 
Also shown are the results of Direct Numerical Simulations \cite{Ishihara2007,Kerr1985}. Note that in \cite{Ishihara2007} more numerical and experimental results are available including the experimental compilation by \cite{Sreeni97}. We chose however those which give the general trend. The DNS results show a slightly increasing trend from $0.4$ to $0.6$ for a Reynolds number going from $10$ to $1000$.   

\subsection{Influence of large-scale forcing}

\begin{figure}
\centering
\setlength{\unitlength}{0.45\textwidth}
\includegraphics[width=1\unitlength]{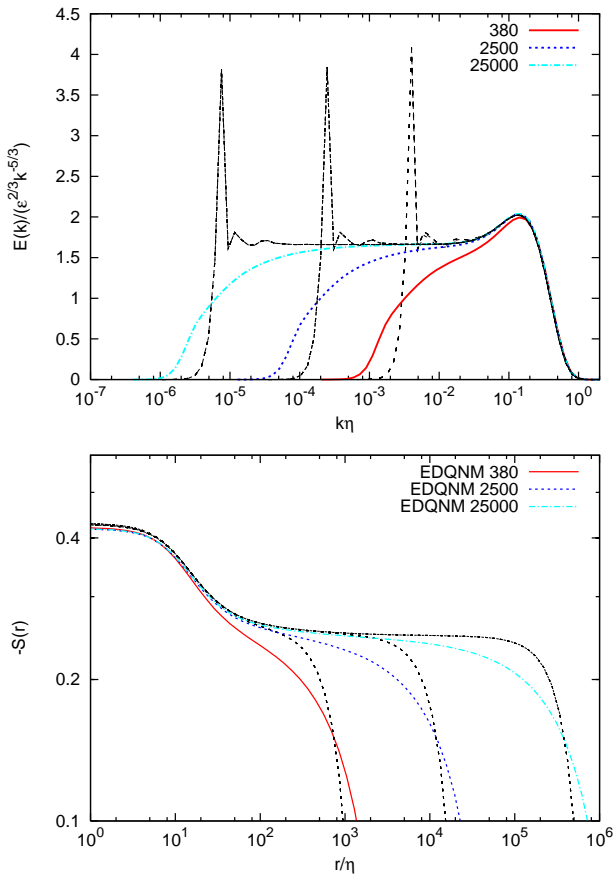}\caption{The influence of a large scale forcing on scaling are illustrated for the energy spectrum and the velocity increment skewness. The black-dotted lines correspond to forced turbulence at the same Reynolds number as the decaying cases considered.
\label{Fig-FD}}
\end{figure}

To conclude this results section we address the influence of a large scale forcing on the scaling of the velocity increment skewness. Indeed, in experiments of nearly isotropic turbulence we often consider a turbulence generated by a grid, advected by a mean velocity. This corresponds in the frame moving with the mean flow, to freely decaying turbulence. Direct numerical simulations of isotropic turbulence are often forced at the large scales in order to obtain a as high as possible Reynolds number. The difference between the two types of turbulent flows is important. For example in \cite{Bos2007-2} it was shown that the normalized dissipation rate is nearly twice as high in decaying turbulence as it is in forced turbulence.  
Also for the appearance of scaling ranges this difference can be important. The difference between decaying and forced turbulence in approaching the asymptotic form of the third-order structure function  was reported in \cite{Kaneda2008}. In figure \ref{Fig-FD} we show how the inertial range scaling of the energy spectrum improves when considering statistically stationary forced turbulence at the same Reynolds number. We observe the appearance of a large peak in the spectrum, corresponding to the forcing. We also observe that the compensated energy spectra display a clear scaling range, already at a Reynolds number of $R_\lambda=380$. The velocity-increment skewness of these forced calculations shows however no clear plateau at this Reynolds number, but its inertial range behavior follows closely the $R_\lambda=2500$ decaying turbulence result.

\section{Conclusion\label{sec5}}

In the present work we computed second and third order structure
functions from EDQNM results. We compared these structure functions
with results from the multifractal formalism. It was shown that in the
near dissipation range the different approaches give very similar
results. It was shown that the appearance of clear scaling ranges is
very slow for the structure functions as was also observed in previous
work \cite{Qian1997,Bos2005}. 

The results for the velocity increment skewness were also compared to experimental results. It was shown that the intermittency 
correction to this quantity given by the multifractal model almost collapsed with the scaling-correction induced by the 
finiteness of the Reynolds number in the EDQNM simulations. In particular at a Reynolds of $R_\lambda=2500$ 
the two corrections almost coincide. This shows that at Reynolds numbers currently achievable in controlled experiments and 
simulations, intermittency corrections to the skewness can not be distinguished from low Reynolds number effects. 
An interesting perspective is to investigate to what extent intermittency corrections to higher-order quantities such as the 
flatness can be distinguished from Reynolds number effects (see e.g. Ref. \cite{ChevillardCastaing2005}). 
This task is within the framework of closure-theory far from trivial and will be left for future work.

\appendix

\section{Relation between second order structure function and the
  energy spectrum}

We will give here a detailed derivation of the relation between second
order structure functions and the kinetic energy spectrum. Parts of
this derivation can be found in different textbooks, but we think it
is useful for the interested reader to give all the details in this work.

Starting from (\ref{eqDLL0}), homogeneity allows to write 
\begin{eqnarray}\label{eqDll0}
D_{LL}(r)&=&2\left(\overline{u_L^2}-\overline{u_Lu'_L} \right)\\
&=&2\frac{r_ir_j}{r^2}\left(\overline{u_iu_j}-\overline{u_iu'_j} \right).
\end{eqnarray}
Using the inverse Fourier transform, and (\ref{eqEk}) we can relate this to the energy spectrum 
\begin{eqnarray}
D_{LL}(r)&=&2\frac{r_ir_j}{r^2}\left(\overline{u_iu_j}-\int\overline{u_i(\bm k)u_j(-\bm k)}e^{i\bm k\cdot\bm r} \ud\bm k \right)\\
&=&\frac{r_ir_j}{r^2}\left(\overline{u_iu_j}-\int \frac{P_{ij}(\bm k)}{4\pi k^2}E(k) e^{i\bm k\cdot\bm r} \ud\bm k \right)
\end{eqnarray}
Defining $\phi$ the angle between $\bm k$ and $\bm r$, we find
\begin{equation}
\frac{r_ir_j}{r^2}P_{ij}(\bm k)=(1-\cos^2 \phi).
\end{equation}
Also, in isotropic turbulence, the Reynolds stress tensor takes the
form,
\begin{equation}
\overline{u_iu_j}=\overline{u^2}\delta_{ij}, ~~~ \overline{u^2}=\frac{2}{3}\int E(k) \ud k,
\end{equation}
so that, introducing conveniently oriented spherical coordinates, we write,
\begin{eqnarray}
D_{LL}(r)
&=&2\overline{u^2}-2\int \frac{(1-\cos^2 \phi)}{4\pi k^2}E(k) e^{i\bm k\cdot\bm r} 2\pi k^2 \sin \phi \ud\phi \ud k \nonumber\\
&=&\int E(k)\left[ \frac{4}{3}-\int (1-\cos^2 \phi)e^{i\bm k\cdot\bm r}\sin \phi \ud\phi \right] \ud k.\nonumber\\
\end{eqnarray}
The integral over $\phi$ can be performed analytically by introducing $\zeta=\cos \phi$ and $x=kr$:
\begin{eqnarray}
&&
\int_0^{\pi}  (1-\cos^2\phi)~\sin\phi~ e^{i kr \cos\phi}\ud\phi\nonumber\\ 
&=&
\int_{-1}^{1} (1-\zeta^2)~  e^{i x \zeta} d\zeta \nonumber\\ 
&=&
\left(1+\frac{\partial^2}{\partial x^2}\right)\int_{-1}^{1}e^{i x \zeta} d\zeta \nonumber\\ 
&=&
\left(1+\frac{\partial^2}{\partial x^2}\right)\frac{e^{i x}-e^{-i x}}{ix}  \nonumber\\ 
&=&
2\left(1+\frac{\partial^2}{\partial x^2}\right)\frac{\sin x}{x}\nonumber\\ 
&=&
4\left(\frac{\sin x}{x^3}-\frac{\cos x}{x^2}\right),
\end{eqnarray}
yielding
\begin{eqnarray}
D_{LL}(r)&=&4\int E(k)\left[ \frac{1}{3}- \frac{\sin (kr) -(kr)\cos (kr)}{(kr)^3} \right]\ud k.
\end{eqnarray}

\section{Relation between third order structure function and the
  energy transfer spectrum}


The Fourier transform of $\overline{u_L{u'_L}^2}$ with respect to $\bm r$ is
\begin{eqnarray}
&&\mathcal FT_{\bm r}\left[\overline{u_L{u'_L}^2}\right]=\\
&=&\overline{u_L(\bm k)u_L^2(-\bm k)}\\
&=&\overline{u_L(\bm k)\iint u_L(\bm p)u_L(\bm q)\delta(-\bm k-\bm p-\bm q)\ud\bm p\ud\bm q}\\
&=&\iint \overline{u_L(\bm k)u_L(\bm p)u_L(\bm q)}\delta(\bm k+\bm p+\bm q)\ud\bm p\ud\bm q\\
&=&\frac{r_ir_jr_m}{r^3}\iint \overline{u_i(\bm k)u_j(\bm p)u_m(\bm q)}\delta(\bm k+\bm p+\bm q)\ud\bm p\ud\bm q\nonumber\\ \\
&=&\frac{r_ir_jr_m}{r^3}T_{ijm}(\bm k)
\end{eqnarray}
Analogously we find
\begin{eqnarray}
\mathcal FT_{\bm r}\left[\overline{u_L^2u'_L}\right]=\frac{r_ir_jr_m}{r^3}T^*_{ijm}(\bm k)
\end{eqnarray}
So that 
\begin{equation}\label{eqDlll1}
D_{LLL}(r)=3\frac{r_ir_jr_m}{r^3}\int \left(T_{ijm}(\bm k)-T^*_{ijm}(\bm k)\right)e^{i\bm k \cdot \bm r}\ud\bm k.
\end{equation}
It is clear from the definitions (\ref{T1}) and (\ref{T2}) that 
$\left(T_{ijm}(\bm k)-T^*_{ijm}(\bm k\right)$
is a third order tensor, symmetric in the indices $j,m$ and solenoidal in the index $i$, so that its general form is
\begin{equation}
T_{ijm}(\bm k)-T^*_{ijm}(\bm k)=\mathcal T(k)P_{ijm}(\bm k)
\end{equation}
after multiplication of both sides by $P_{ijm}(\bm k)$ one finds 
\begin{equation}
\mathcal T(k)=\frac{P_{ijm}(\bm k)}{4k^2}\left(T_{ijm}(\bm k)-T^*_{ijm}(\bm k)\right)=\frac{T(k)}{4i\pi k^4}.
\end{equation}
We substitute this in (\ref{eqDlll1}),
\begin{equation}\label{eqDlll2}
D_{LLL}(r)=3\frac{r_ir_jr_m}{r^3}\int \frac{T(k)}{4i\pi k^4}P_{ijm}(\bm k)e^{i\bm k \cdot \bm r}\ud\bm k.
\end{equation}
Defining $\phi$ the angle between $\bm k$ and $\bm r$, we find that
\begin{equation}
\frac{r_ir_jr_m}{r^3}P_{ijm}(\bm k)=2k\cos \phi(1-\cos^2 \phi)
\end{equation}
and thus 
\begin{equation}\label{eqDlll3}
D_{LLL}(r)=6\int \cos \phi(1-\cos^2\phi) \frac{T(k)}{4i\pi k^3}e^{i\bm k \cdot \bm r}\ud\bm k.
\end{equation}
Introducing again  conveniently oriented spherical coordinates, we write this as 
\begin{eqnarray}\label{eqDlll4}
D_{LLL}(r)&=&6\int_0^{\infty}\int_0^\pi \cos \phi(1-\cos^2\phi) \frac{T(k)}{4i\pi k^3}e^{i\bm k \cdot \bm r}\times \nonumber\\
&&2\pi k^2 \sin\phi \ud\phi \ud k\\
&=& -3i\int_0^{\infty} \frac{T(k)}{k} \int_0^\pi \cos \phi(1-\cos^2\phi)e^{i\bm k \cdot \bm r}  \times\nonumber\\
&&\sin\phi \ud\phi \ud k.
\end{eqnarray}
As for $D_{LL}(r)$, the integral over $\phi$ can be performed analytically by introducing $\zeta=\cos \phi$ and $x=kr$:
\begin{eqnarray}
&&
\int_0^{\pi}  cos\phi~ (1-\cos^2\phi)~\sin\phi~ e^{i kr \cos\phi}\ud\phi\nonumber\\ 
&=&
\int_{-1}^{1} \zeta(1-\zeta^2)~  e^{i x \zeta} d\zeta \nonumber\\ 
&=&
-2i\left(\frac{2\sin x}{x^2}+\frac{6\cos x}{x^3}-\frac{6\sin x}{x^4} \right),
\end{eqnarray}
yielding
\begin{eqnarray}
D_{LLL}(r)&=& 12r\int_0^{\infty} T(k) \frac{3 \left(\sin kr-kr\cos kr\right)-(kr)^2\sin kr}{(kr)^5} \ud k\nonumber\\
\end{eqnarray}

\section{The multifractal description}\label{AppMF}

The multifractal formalism can be seen as a probabilistic interpretation of the  averaged behavior of velocity structure functions. More precisely, for a scale $r$ in the inertial range, using both the standard arguments of 
the multifractal formalism \cite{FrischBook} and the probabilistic formulation of Castaing \cite{CasGag90}, 
the velocity increment $\delta u_L(x,r) = u_L(x+r)-u_L(x)$ can be represented by the product of two independent 
random variables,  $\delta u_L(x,r) = \beta_r \xi$, with $\xi$  a zero average Gaussian random variable of variance 
$\sigma^2=\langle [\delta u_L (x,L_0)]^2 \rangle$, where $L_0$ is the integral length scale, 
and a stochastic variance $\beta_r = \left(\frac{r}{L_0}\right)^h$ where the exponent $h$ fluctuates itself according to the 
law $\mathcal P_r(h) \propto \left(\frac{r}{L_0}\right)^{1-\mathcal D(h)}$. This gives a complete one-point probabilistic 
description (including structure functions and probability density functions) of the velocity increments in the inertial range 
given by an empirical function $\mathcal D(h)$. This function is
both scale and Reynolds number independent, and is called the singularity spectrum in the inviscid limit. 
Paladin and Vulpiani \cite{PalVul87} and Nelkin \cite{Nel90} then proposed a natural extension to the 
dissipative scales and the respective description of the velocity
gradients. This adds to the description a Reynolds dependence 
through the fluctuating nature of the dissipative scale $\eta(h) = L_0(\mathcal R_e/\mathcal
R^*)^{-1/(h+1)}$, where $\mathcal R_e = \sigma L_0/\nu$ and $\mathcal R^*=52$ a universal constant related to the 
Kolmogorov constant \cite{Chevillard2006-2}. The relation between this
Reynolds number and the Taylor-scale Reynolds number is
\begin{equation}
\mathcal R_e = \frac{4}{\mathcal R^*}R_\lambda^2.
\end{equation}
Let us remark that in a
K41 framework the variable $h=1/3$ is unique and does 
not fluctuate and one recovers the classical Kolmogorov prediction $\eta_K=\eta(h=1/3) = L_0(\mathcal R_e/\mathcal
R^*)^{-3/4}$. 

Meneveau \cite{Men96} proposed an elegant interpolation formula between the inertial range and the 
far dissipative range. Following these works reference
[\onlinecite{Chevillard2006-2}] proposed a probabilistic 
formulation of velocity increments that covers the entire range of
scales. The expressions for  $\beta_r$ and $\mathcal P_r(h)$ in
this description are
\begin{equation}\label{eq:betah}
\beta_r  =
\frac{\left(\frac{r}{L_0}\right)^h}{\left[
1+\left(\frac{r}{\eta(h)}\right)^{-2}\right]^{(1-h)/2}} \mbox{
,}
\end{equation}
and
\begin{equation}\label{eq:phmeneeul}
\mathcal P_r(h)
=\frac{1}{\mathcal Z(r)}
\frac{\left(\frac{r}{L_0}\right)^{1-\mathcal D(h)}}{\left[
1+\left(\frac{r}{\eta(h)}\right)^{-2}\right]^{(\mathcal
D(h)-1)/2}}\mbox{ ,}
\end{equation}
where $\mathcal Z (r)$ is a normalization factor such that
$\int_{h_{\min}}^{h_{\max}} \mathcal P_r(h) dh= 1$. We will take $h_{\min}=0$ and $h_{\max}=1$. Given the parameters 
of the flow, namely $L_0$, the large scale variance $\sigma^2$ and $\mathcal R_e$, this description requires one additional free parameter 
$\mathcal R^*$ and a parameter function $\mathcal D(h)$ that can be measured from empirical data. 
We will take $\mathcal R^* = 52$ and a parabolic approximation for the singularity spectrum 
$\mathcal D(h) = 1-\frac{(h-c_1)^2}{2c_2}$. The so-called intermittency coefficient has been estimated from data to be 
$c_2=0.025$ \cite{Chevillard2006-2}. The remaining parameter $c_1$ is chosen such that, in the inertial range, 
the third order structure function $\langle |\delta u_L|^3\rangle$ is proportional to the scale $r$. This gives 
$c_1=\frac{1}{3}+\frac{3}{2}c_2$. 

The proposed description has been shown to accurately describe the symmetric part of the 
velocity increments probability density functions 
and even order structure functions. In particular, using Eqs. (\ref{eq:betah}) and (\ref{eq:phmeneeul}), even order structure functions 
are given by the following integral
\begin{equation}
\langle [\delta u_L(x,r)]^{2q}\rangle = \langle \xi^{2q}\rangle\int_{h_{\min}}^{h_{\max}}\beta_r^{2q} \mathcal P_r(h) dh\mbox{ ,}
\end{equation}
with $\langle \xi^{2q} \rangle = \sigma^{2q}\frac{(2q)!}{q!2^q}$. 
Furthermore, one can show \cite{Chevillard2006-2} that the mean dissipation
\begin{equation}
\langle \epsilon\rangle = 15\nu \langle (\partial _x u )^2\rangle
= 15\nu \lim_{r\rightarrow 0} \frac{\langle (\delta u_L)^{2q}\rangle}{r^2}\approx \frac{\sigma^3}{L_0}\frac{15}{\mathcal R^*}
\end{equation} 
is independent on the Reynolds number. Since the Gaussian noise $\xi$ is independent on 
the fluctuating exponent $h$, odd order structure functions vanish. Some modifications of the noise $\xi$ have been proposed in 
Refs. \cite{CasGag90} and \cite{Chevillard2006-2} in order to take
into account a non-zero skewness. Nevertheless, without any 
additional free parameters, the Karman-Howarth-Kolmogorov equation
\begin{equation}\label{eq:KarHow}
\langle (\delta u_L)^3\rangle = -\frac{4}{5}\langle \epsilon
\rangle r + 6\nu \frac{d\langle (\delta
u_L)^2\rangle}{dr}\mbox{ ,}
\end{equation}
gives a prediction for the third order moment $\langle (\delta u_L)^3\rangle$, and thus the skewness, knowing only the second 
order one. This study was carried out and compared against experimental data in Ref. \cite{Chevillard2006-2}. 
As a final remark, we would like to add that such an approach allows to give a prediction for the third order moment of velocity 
gradients, i.e.
\begin{align}\label{eq:pred1}
\langle (\partial_xu)^3\rangle &= -\frac{6\nu\sigma^2}{L^4} \left[
\frac{2}{\mathcal Z(0)}
\int_{h_{\min}}^{h_{\max}}\right. \notag \\
&\left.\left[2h-1-\mathcal D(h) \right]\left(
\frac{\eta(h)}{L}\right)^{2(h-2)+1-\mathcal D(h)}dh + \mathcal F
\right]\mbox{ ,}
\end{align}
where $\mathcal F$ is a negligible additive term, coming from the
Taylor's development of the normalization factor $\mathcal
Z(r)$, and given in \cite{Chevillard2006-2}. Using Eq. (\ref{eq:pred1}), a steepest-descent calculation shows that
the skewness of the derivatives behaves as a power law of the Reynolds number, i.e. $-S(0) \sim \mathcal R_e^{\chi-1}$ with
\begin{align}
\chi &=  \min_h \left[ -\frac{2(h-2)+1-\mathcal D(h)}{h+1}\right]\notag \\
&-\frac{3}{2}\min_h \left[ -\frac{2(h-1)+1-\mathcal D(h)}{h+1}\right] \mbox{ .}
\end{align}
Using a quadratic approximation for the parameter function $\mathcal D$ and $c_2 = 0.025$, one gets
\begin{equation}\label{eq:predSkewGradLN}
 -S(0) \sim \mathcal R_e^{0.067}\mbox{ .}
\end{equation}
These result were already obtained by Nelkin \cite{Nel90} using a different, although related, approach, based on 
the asymptotically exact relationship $\langle (\partial_xu)^3\rangle = -2\nu \langle (\partial_x^2u)^2\rangle$.  

Finally, we need to discuss the behaviour of the multifractal description at scales of the order of the integral length scale $L_0$. 
At this stage, the proposed formalism (Eqs. (\ref{eq:betah}) and (\ref{eq:phmeneeul})) is valid only in the limit $r\ll L_0$, 
such that
the integrals correspond to their steepest-descent values. The
multifractal description describes hereby an asymptotic state in which the
influence of the decay of the turbulence or the energy-input mechanism
is not taken into account. The inertial range starts directly at the
integral scale. Second order structure functions do therefore not
smoothly tend to a constant value around $L_0$ and the third-order
structure function does not tend to zero. The present work is not
devoted to an extension of the description to take into account
explicitly the large scales. However, in order to compute the energy
and transfer spectra from structure functions through the relations
(\ref{DtoE}) and (\ref{DtoT}) we need a smooth
behavior around the large scales for the integrals to converge. For
this reason we propose an \emph{ad-hoc} modification of the
multifractal description. Firstly we use $\beta_r = 1$ (i.e., independent on $h$ and non fluctuating) 
and $\mathcal P(h)$ uniform for $r \geq L$. This is equivalent to a Gaussian modeling of the velocity increments, 
with a scale independent variance. Unfortunately, this  description is not continuous (because of the dissipative corrections). 
Moreover, one has to extend the validity of the 
Karman-Howarth-Kolmogorov equation (\ref{eq:KarHow}) itself in order to get a realistic behaviour of the third order moment at 
large scales. In order to get a continous and differentiable in scale
description of the velocity increments, we propose to replace the
scale $r$ entering Eqs. (\ref{eq:betah}) and (\ref{eq:phmeneeul})
by the ersatz $r'=L_0~ \mbox{tanh}(r/L_0)$. The hyperbolic tangent
allows a smooth transition to $r'=r$ for $r\ll L_0$ and
$r'=L_0$ for $r\gg L_0$. Still, this does not fix the unrealistic
behavior of Eq. (\ref{eq:KarHow}) at large scales. Thus, the
prediction of $\langle [\delta u(x,r')]^3\rangle$ obtained from  Eq.
(\ref{eq:KarHow}) using the ersatz $r'$ is furthermore multiplied by
a large-scale cutoff of the form $\exp(-r'^2/(2L_0^2))$. We stress again that
this approach is completely \emph{ad-hoc} and only used in order to
allow computation of wave-number spectra.


\end{document}